\documentstyle[12pt,twoside]{article}
\newcommand{\nc}{\newcommand}
\nc{\Real}{{\rm Re}} \nc{\Imag}{{\rm Im}} \nc{\Lg}{{\cal {L}}}
\nc{\al}{\alpha} \nc{\be}{\beta} \nc{\ga}{\gamma} \nc{\de}{\delta}
\nc{\vare}{\varepsilon} \nc{\te}{\theta}  \nc{\la}{\lambda}
\nc{\sig}{\sigma} \nc{\var}{\varphi} \nc{\om}{\omega}
\nc{\De}{\Delta} \nc{\Ga}{\Gamma} \nc{\La}{\Lambda} \nc{\Sig}{\Sigma} \nc{\Om}{\Omega}
\nc{\bb}{\begin{equation}} \nc{\ee}{\end{equation}} \nc{\bega}{\begin{eqnarray}}
\nc{\eea}{\end{eqnarray}} \nc{\begea}{\begin{eqnarray*}} \nc{\eqea}{\end{eqnarray*}}
\nc{\pa}{\partial}     \nc{\lan}{\langle}     \nc{\ran}{\rangle}
\nc{\pan}{{\pa_{\nu}}}   \nc{\pam}{{\pa_{\mu}}}
\nc{\panb}{{\frac{\pa}{\pa {x^{'}}^{\nu}}}}
\nc{\pamb}{{\frac{\pa}{\pa {x^{'}}^{\mu}}}}
\nc{\longr}{\longrightarrow} \nc{\Longr}{\Longrightarrow}
\nc{\Longlr}{\Longleftrightarrow} \nc{\leria}{\leftrightarrow} \nc{\ria}{\rightarrow}
\nc{\C}{I\!\!\!C} \nc{\1}{1\!\!1} \nc{\R}{I\!\!R}
\nc{\ps}{/\!\!\!p} \nc{\pis}{/\!\!\!\pi} \nc{\ii}{\`{\i \ }}
\nc{\as}{/\!\!\!a} \nc{\bs}{/\!\!\!b}
\nc{\Orm}{{\rm O}\!\!\!\!\!{\rm O}}
\nc{\h}{\hspace*{0.5 cm}} \nc{\hs}{\hspace*} \nc{\vs}{\vspace*}
\nc{\widt}{\widetilde} \nc{\widh}{\widehat}
\nc{\ov}{\overline} \nc{\und}{\underline} \nc{\cent}{\centerline}
\nc{\dis}{\displaystyle} \nc{\ug}{\; = \;} \nc{\dg}{\dagger} \nc{\st}{\star}
\nc{\xbf}{\mbox{\boldmath $x$}} \nc{\pabf}{\mbox{\boldmath $\partial$}}
\nc{\vbf}{\mbox{\boldmath $v$}} \nc{\imp}{\mbox{\boldmath $p$}}
\nc{\wbf}{\mbox{\boldmath $w$}} \nc{\sbf}{\mbox{\boldmath $s$}} \nc{\jbf}{\mbox{\boldmath $j$}}
\nc{\Psirm}{\mbox{\boldmath ${\rm \Psi}$}}
\nc{\Psibrm}{\mbox{\boldmath ${\ov {\rm \Psi}}$}}
\nc{\Psidrm}{\mbox{\boldmath ${\rm {{\Psi}^{\dg}}}$}}
\nc{\Imp}{\mbox{\boldmath ${P}$}} \nc{\Tbf}{\mbox{\boldmath ${T}$}}
\nc{\Qbf}{\mbox{\boldmath ${Q}$}} \nc{\Hbf}{\mbox{\boldmath ${H}$}}
\nc{\Jbf}{\mbox{\boldmath ${J}$}} \nc{\Vbf}{\mbox{\boldmath ${V}$}}
\nc{\Sbf}{\mbox{\boldmath ${S}$}} \nc{\Lgbf}{\mbox{\boldmath $\Lg$}}
\nc{\albf}{\mbox{\boldmath $\al$}} \nc{\gabf}{\mbox{\boldmath $\ga$}}
\nc{\sigbf}{\mbox{\boldmath $\sigma$}} \nc{\vecna}{\mbox{\boldmath $\nabla$}}
\nc{\Sigvec}{{\vec {\Sigma}}}
\nc{\ao}{{\widh a}}    \nc{\bo}{{\widh b}}    \nc{\No}{{\widh N}}
\nc{\aod}{{\ao}^{\dg}}       \nc{\bod}{{\bo}^{\dg}}
\nc{\vbo}{{\widh{\vbf}}}
\nc{\xbo}{{\widh{\xbf}}}  \nc{\xdo}{{\widh{{\dot {x}}}}} \nc{\Ho}{{\widh H}}
\nc{\Po}{{\widh P}} \nc{\Co}{{\widh C}} \nc{\Qo}{{\widh Q}} \nc{\To}{{\widh T}}
\nc{\lao}{{\widh \la}} \nc{\po}{{\widh p}} \nc{\Io}{{\widh I}} \nc{\Lo}{{\widh L}}
\nc{\ptr}{{\stackrel{\leftarrow}{\partial}}}
\nc{\xbdo}{{\widh{{\dot {\xbf}}}}}
\nc{\sbo}{{\widh{\sbf}}}   \nc{\impo}{\widh{\imp}}
\nc{\psib}{\ov {\psi}}     \nc{\psid}{{\psi}^{\dg}}
\nc{\psis}{\psi^{\st}} \nc{\psit}{\widt{\psi}}
\nc{\erm}{{\rm e}}  \nc{\um}{{1\over 2}}  \nc{\ub}{{\ov u}}
\nc{\ddox}{\ddot{x}} \nc{\dox}{\dot{x}} \nc{\qq}{\qquad \qquad}
\nc{\gm}{\ga^{\mu}} \nc{\gc}{\ga^{5}} \nc{\go}{\ga^{0}} \nc{\alc}{\al^{5}}
\nc{\lf}{{\rm L}}  \nc{\rg}{{\rm R}} \nc{\rd}{{\rm d}}
\nc{\npt}{{1\over{\sqrt{2kV}}}}  \nc{\nbr}{{1\over{\sqrt{2\vare V}}}}
\nc{\teta}{{\theta_\la}} \nc{\tetad}{{{\theta_\la}^\dg}}
\nc{\tetam}{{\theta_{-\la}}} \nc{\tetamd}{{{\theta_{-\la}}^\dg}}
\nc{\xilm}{{\xi_{-\la}}} \nc{\xidlm}{{{\xi_{-\la}}^\dg}}
\nc{\xil}{{\xi_\la}} \nc{\xidl}{{{\xi_\la}^\dg}} \nc{\vc}{\vbf_{\rm cl}}
\nc{\Apl}{A_{p\la}}  \nc{\Bpl}{B_{p\la}}  \nc{\ap}{a_{p\la}} \nc{\upl}{u_{p\la}}
\nc{\apb}{a_{p'\lb}} \nc{\uplb}{u_{p'\lb}}  \nc{\bp}{b_{p\la}} \nc{\vpl}{v_{p\la}}
\nc{\bpb}{b_{p'\lb}} \nc{\vplb}{v_{p'\lb}}   \nc{\varb}{\vare^{'}}
\nc{\imb}{\imp^{'}}  \nc{\lb}{{\la}^{'}}  \nc{\intp}{\int\rd^3p}
\nc{\Bt}{{\widetilde{B}}}  \nc{\rB}{{\rm B}}  \nc{\rT}{{\rm T}}
\nc{\pmut}{\widetilde{p}} \nc{\vbm}{{\ov {\vbf}}} \nc{\sbm}{{\ov {\sbf}}}
\nc{\vmm}{{{\ov {v}}^\mu}} \nc{\smm}{{{\ov {s}}^\mu}} \nc{\rinv}{{\rho^{-1}}}
\nc{\vm}{{\ov {v}}} \nc{\sm}{{\ov {s}}} \nc{\som}{{{\ov {s}}^0}}
\nc{\dV}{\rd V}  \nc{\ja}{j_{{\rm A}}}  \nc{\Sb}{{\bar{S}}}
\nc{\bi}{\bibitem} \nc{\ia}{\'{\i}}

\renewcommand{\thefootnote}{\fnsymbol{footnote}}

\begin{document}

\small

\cent{{\em Published in} Int.\,J.\,Mod.\,Phys.\,{\bf A28}\,(1997)\,5103}

\

\cent{\bf SLOWER-THAN-LIGHT SPIN-$\um$ PARTICLES}
\cent{\bf ENDOWED WITH NEGATIVE MASS SQUARED$^{(\dg)}$}
\footnotetext{$^{(\dg)}$ Work partially supported by INFN}
\vs{0.8cm}
\begin{center}
{Giovanni Salesi

\vs{0.4cm}

{\em Dipartimento di Fisica, Universit\`a Statale di Catania, I-95129--Catania,
Italy; \ and

Istituto Nazionale di Fisica Nucleare---Sezione di Catania, Catania, Italy;

e-mail: {\rm Salesi@ct.infn.it}}}
\end{center}
\vs{0.5cm}
\begin{abstract}
Extending in a straightforward way the standard Dirac theory, we study a quantum
mechanical wave-equation describing free spinning particles ---which we propose
to call {\em Pseudotachyons} (PT's)--- which behave like tachyons in the
momentum space ($p^2=-m^2$), but like subluminal particles ($v<c$) in the
ordinary space. This is allowed since, as it happens in every quantum theory
for spin-$\um$ particles, the momentum operator, $-i\vecna,$ (that is conserved)
and the velocity operator $\albf$ (that is not) are independent operators,
which refer to independent quantities: $\impo\neq m\vbo$. \ As a consequence,
at variance with ordinary Dirac particles, for PT's the average velocity
$\vbm\equiv\lan\psid\albf\psi\ran/\lan\psid\psi\ran$ is not equal to the
classical velocity $\vc=\imp/\vare$, but actually to the velocity ``dual'' of
$\vc$: $\vare\imp/\imp^2$. Being reciprocal of $|\vc|$, the speed of PT's is
therefore smaller than the speed of light. Since a lot of experimental
data seems to involve a negative mass squared for neutrinos, we suggest that
these particles might be PT's, travelling, because of their very small mass,
at subluminal speeds very close to $c$. The present theory is shown to be
separately invariant under the $C$, $P$, $T$ transformations; the covariance
under Lorentz transformations is also proved. Furthermore, we derive the
kinematical constraints linking 4-impulse, 4-velocity and 4-polarization of
free PT's.

\

\noindent PACS numbers: 03.65.Pm; 12.39.Fe; 14.60.St
\end{abstract}

\section{Dirac-like equation for spin-$\um$ particles endowed with negative
mass squared}

One of the simplest Dirac-like lagrangians, at the first order in the scalar
bilinear $\psib\psi$, hermitian and relativistically invariant,
which is able to describe (as it will be shown) spin-$\um$ free particles
endowed with a negative mass squared,
is the following:
\bb
\Lg \ug i\,\psib\gc\gm\pa_{\mu}\,\psi - m\,\psib\psi
\ee
[as usual, we shall hereafter assume: $\hbar=c=1; \ \psib\equiv\psid\go; \
\gc\equiv i\go\ga^1\ga^2\ga^3$, with \ ${(\gc)}^2=\1$ and \ ${(\gc)}^{\dg}=\gc$],
whilst the ordinary Dirac lagrangian (describing the so-called {\em
bradyons} endowed with positive 4-impulse squared) writes
\bb
\Lg \ug i\psib\gm\pa_{\mu}\psi - m\,\psib\psi\,.
\ee
For reasons which will be clarified in the fourth section henceforth we shall
call {\em Pseudotachyons} (PT's) the spinning particles with negative 4-impulse
described by eq.~$\!$(1).
This lagrangian, firstly introduced by Tanaka$^{[1]}$, has been re-proposed in
order to describe ``tachyonic neutrinos'' about ten years ago by Chodos
{\em et al.}$^{[2]}$. In this paper we study, within a new formal approach, the
plane wave solutions of the PT Dirac equation entailed by this lagrangian. In so
doing we shall deduce out peculiar physical results.
Actually, at variance with the previous literature on spinning quantum particles
endowed with $p^2<0,^{[3-7]}$ we shall find that {\em such particles are
slower-than-light particles}. In the last section we shall prospect the possible
observation of PT's as ordinary neutrinos.

Before proceeding further let us advise that, for the present work, {\em
no extension of special relativity is demanded} and that {\em only standard
Lorentz transformations with subluminal boosts of the reference frames are
possible.}
In such a way timelike 4-impulses transform into
timelike 4-impulses and spacelike 4-impulses into
spacelike 4-impulses, without any change of the sign of the mass squared $p^2$.
As a consequence, we have {\em two distinct theories}, one generated
by lagrangian (1), and one by lagrangian (2),
referring to two really different types of particles. These theories entail
{\em in every reference frame} opposed sign for $p^2$, even if, as it will be
shown in the last section, both types of particles appear to travel at subluminal speeds.
Among the most noticeable differences there is, e.g., {\em the non-existence,
for PT's,
of the center-of-mass frame}, where $\imp=0$, because we always have
$|\imp|\geq m$ in every frame. In fact, only through a ``superluminal'' Lorentz
transformation with a boost larger than $c$, the particle momentum could vanish
with respect to a given reference frame. Nevertheless, we shall see below that
{\em there exists the ``quiet'' frame} in which, not the momentum, but the
velocity vanishes.

\

\noindent Taking the variation with respect to the ``lagrangian coordinate'' $\psib$ we
obtain the {\em PT Dirac-like equation}:
\bb
(i\gc\gm\pa_{\mu}\, - \,m)\,\psi \ug 0\,,
\ee
in which the kinetic term differs from the one of the standard Dirac equation
\ $(i\gm\pa_{\mu}\, - \,m)\,\psi=0$\,.
Notice the analytic continuity of equation (3) with the Dirac equation:
the ``right limit'' of the PT equation for $v\longr c^+$ coincides with the
``left limit'' for $v\longr c^-$ of the  bradyonic equation, because for
$m\longr0$ the two equations entail the same solutions (see also the next section).

Usually$^{[3-7]}$ the constitutive relation of tachyonic kinematics, \ $p^2=-m^2$, \
is found
by making recourse to non-hermitian terms in the lagrangian which involve an
{\em imaginary mass}. On the contrary, by means of lagrangian (1), this
constraint is obtained
simply as a direct consequence of the {\em non-commutativity}
of the hermitian matrix $\gc$ with the matrices $\gm$.
In fact, multiplying eq.~$\!$(3) left for $(i\gc\gm\pa_{\mu}\, + \,m)$,
\bb
(i\gc\gm\pa_{\mu}\, + \,m)\,(i\gc\gm\pa_{\mu}\, - \,m)\,\psi \ug 0\,,
\ee
for the anticommutativity of $\gm$ with $\gc$, which implies $(\gc\gm)^2=-\1$, we
get:
\bb
(\Box\, - \,m^2)\,\psi \ug 0\,,
\ee
namely, the so-called {\em Klein--Gordon equation for tachyons}.$^{[8,9]}$

For the ``hermitian adjoint'' wave-function $\psib$ the following equation holds
(where $i\ptr$ indicates the transposed
4-impulse operator which is ``left-acting'')
\bb
\psib\,(i\gc\gm\ptr_\mu\, + \,m) \ug 0\,.
\ee
Expliciting in (3) the product $\gc\gm\pa_{\mu}$, and left multiplying by
$\go\gc$, namely
\bb
(i\pa_t - i\albf\cdot\vecna)\,\psi \ug m\go\gc\psi
\ee
(where, as usual, $\albf\equiv\go\gabf$), we obtain the hamiltonian operator for
free PT's:
\bb
\Ho \ug -i\albf\cdot\vecna + m\alc\,.
\ee
In the last expression we have renamed $\alc$ the $\go\gc$ matrix, with
$({\alc})^2 = -\1$.
\ Obviously, applying to $\psi$ the hamiltonian squared $\Ho^2$ we obtain:
\bb
\Ho^2\,\psi = -\pa_t^2\psi \ug (-\vecna^2 - m^2)\,\psi\,,
\ee
that is, the above Klein--Gordon equation for tachyons.

Like in the bradyonic theory, also for PT's the spin component along the
momentum direction, i.e. the helicity,
conserves. In fact, as is easily proved,
the PT hamiltonian (8) commutes with the spin$\times$momentum operator:
\ $[\Ho, \;\um\Sigvec\cdot\impo] \equiv [\Ho, \;\um\albf\gc\cdot\impo] = 0$.

\

\

\section{Plane waves}

Let us write the general plane wave in the typical form met in the standard
spinning particles theory:
\bb
\psi \ug N_{p}\,w_{\pm p}\,\erm^{\mp ipx}\,,
\ee
where $N_p$
is a suitable normalization factor; quantity $w_{\pm p}$ denotes, as usual, the
four-component
``Dirac bispinor'', that is a $p$-function; \ $px
\equiv \vare t - \imp\cdot\xbf$; \ the superior sign refers to the
``positive-energy states'' for which $p^0=+\vare$ (with
$\vare=\sqrt{\imp^2+m^2}$ for bradyons and $\sqrt{\imp^2-m^2}$ for PT's)
and the momentum is $\imp$,
whilst the inferior sign refers to the ``negative energy states''
for which
$p^0=-\vare$ and the momentum is $-\imp$.
As is known, the two superior components and the two inferior
ones in the amplitudes $w_{\pm p}$ are linked through relations whose particular form
depends on the chosen representation for the matrices $\gm$.
We shall work in the so-called {\it standard} representation; for the expressions
of the solutions of the PT Dirac equation in {\em spinorial} or {\em Weyl's}
representation see the Appendix.
In standard representation the matrices $\gm$, $\gc$ are the following:
\bb
\go \equiv \left(
\begin{array}{cc}
\1   & \ \Orm\\
\Orm & -  \1\\
\end{array}
\right)
\qquad
\gabf \equiv \left(
\begin{array}{cc}
\ \Orm     & \sigbf\\
- \sigbf   & \Orm\\
\end{array}
\right)
\qquad
\gc \equiv \left(
\begin{array}{cc}
\Orm &  \1\\
\1   &  \Orm \\
\end{array}
\right)\,,
\ee
where $\1$ is the identity $2\times 2$ matrix, \ $\Orm$ is the null $2\times2$
matrix, \ the components of $\sigbf \equiv (\, \sig_1;\, \sig_2;\, \sig_3)$ are
the usual Pauli matrices.
In this representation the wave amplitude can be written as follows:
\bb
w \equiv
\left(\begin{array}{cc}\varphi\\ \chi\end{array}\right)\,,
\ee
where $\chi$ and $\varphi$ are (two-component)
{\em Pauli spinors}.

Let us now choose the normalization factor in eq.~$\!$(10) as follows:
\bb
\psi \ug \npt\,w_{\pm p}\,\erm^{\mp ipx}\,,
\ee
where $k\equiv|\imp|$.
In the fourth section we shall see that, for PT's, the factor $\npt$ is
a Lorentz invariant quantity, so that $w$ really transforms as a Dirac bispinor.
For such a plane wave, the Klein--Gordon equation (5) leads to the relation which defines the
tachyonic kinematics
\bb
\vare^2 \ug \imp^2 - m^2\,;
\ee
while the PT Dirac equation (3) leads to the following matrix equations for
the bispinorial amplitudes
(with $u_p\equiv w_{p}; \ v_p\equiv w_{-p}; \ \ps \equiv p_{\mu}\ga^{\mu}$):
\bb
(\ps - m\gc)\,u_p \ug 0\,,
\ee
\bb
(\ps + m\gc)\,v_p \ug 0\,.
\ee
In standard representation, from eq.~$\!$(15) we obtain
\bb
{\dis\left\{\begin{array}{l}

\\

\ \vare\,\varphi - (\imp\cdot\sigbf)\,\chi \ug m\,\chi\\

\\

- \vare\,\chi + (\imp\cdot\sigbf)\,\varphi \ug m\,\varphi\,,\\

\end{array}\right.}
\ee
and from eq.~$\!$(16):
\bb
{\dis\left\{\begin{array}{l}

\\

- \vare\,\varphi + (\imp\cdot\sigbf)\,\chi \ug m\,\chi\\

\\

\ \vare\,\chi - (\imp\cdot\sigbf)\,\varphi \ug m\,\varphi\,.\\

\end{array}\right.}
\ee
From system (17) we deduce explictly the positive-energy amplitudes and from
system (18) the negative-energy ones:
\bb
u_p \ug
\left(\begin{array}{cc}
\varphi\\
\ \\
\frac{\imp\cdot\sigbf - m}{\vare}\,\varphi
\end{array}\right)\,,
\qquad\qquad
v_p \ug
\left(\begin{array}{cc}
\frac{\imp\cdot\sigbf-m}{\vare}\,\chi\\
\ \\
\chi
\end{array}\right)\,.
\ee
We have expressed $u_p$ as a function of $\varphi$ rather than of $\chi$, and
$v_p$ as a function of $\chi$ rather than of $\varphi$, for a mere formal
analogy with the bradyonic solutions (see below). This choice is really convenient only for
the Dirac amplitudes because in the center-of-mass frame (or, equivalently, in
the non-relativistic approximation, $\imp\longr0$) $\chi\longr0$ in $u_p$
and $\varphi\longr0$ in $v_p$ [cf. eqs.~$\!$(30) with $k=0$].

Once arbitrarily chosen the $\varphi$ ($\chi$) spinor, the other one,
$\chi$ ($\varphi$), comes out to be univocally determined. The two
degrees of freedom inherent to the choice of the
spinor are to be correlated
to the spin orientation, and then to the two available
helicity states. Thus, for every momentum $\imp$ we shall have two distinct helicity
eigenstates $\psi_{p\la}$. Henceforth $\la$ is defined as twice the helicity, \
$\la \equiv 2\,{{\imp\cdot\sbf}\over{|\imp|}}$\ , but,
for convenience, also $\la$ itself will be called helicity.
\ Let us impose that $\upl$ and $\vpl$ be eigenstates of the operator
$\lao \equiv {{\imp\cdot\Sigvec}\over{|\imp|}}$, where the Dirac 4$\times$4 spin
operator $\Sigvec\equiv\albf\gc$ writes:
\bb
\Sigvec \equiv \left(
\begin{array}{cc}
\sigbf  &  \Orm\\
\Orm    &  \sigbf\\
\end{array}
\right)\,.
\ee
Let us also require the normalization ``to one particle in the 3-volume $V$''
(see the last section), \ $\psid\psi=1/V$, which implies, as it is soon verified,
the normalization
\bb
\upl^\dg \upl=\vpl^\dg \vpl=2k\,.
\ee
From eqs.~$\!$(19), exploiting also the equality
$\vare=\sqrt{(k+m\la)(k-m\la)}$, we have for the helicity eigenfunctions:
\bb
\upl = \sqrt{k+m\la}
\left(\begin{array}{cc}
\teta\\
\ \\
{{\la\vare}\over{k+m\la}}\,\teta
\end{array}\right)\,,
\qquad\qquad
\vpl = \sqrt{k-m\la}
\left(\begin{array}{cc}
{{-\la\vare}\over{k-m\la}}\,\tetam\\
\ \\
\tetam
\end{array}\right)\,.
\ee
In these expressions the spinors $\teta$, $\tetam$ are
2-component
column spinors,
eigenfunctions of $\lao$,
\bb
\lao\,\teta \ug \la\,\teta\,, \qquad \qquad \qquad \lao\,\tetam \ug -\la\,\tetam\,,
\ee
and normalized to unity: $\theta^\dg_{\lb}\teta=\de_{\lb\la}$.
Following the current notation, the amplitude $\vpl$ relates to a state endowed with
helicity $-\la$. This choice
turns out to be convenient for
the so-called
``reinterpretation'' of the negative-energy states,$^{[10]}$ after which
we shall really observe a 4-impulse $+p$ and a helicity $+\la$.

\

\noindent Let us now write down, for a direct comparison, the plane wave solutions of
the standard bradyonic Dirac equation \ $(i\gm\pa_{\mu}\, - \,m)\,\psi=0$\,.
We choose the usual normalization factor $\nbr$ which, for bradyons, constitutes
a Lorentz scalar:
\bb
\psi \ug \nbr\,w_{\pm p}\,\erm^{\mp ipx}\,.
\ee
For such a plane wave the Dirac equation leads to the following matrix equations
for the bispinorial amplitudes:
\bb
(\ps - m)\,u_p \ug 0\,,
\ee
\bb
(\ps + m)\,v_p \ug 0\,.
\ee
In standard representation, from the former of the above equations we get
\bb
{\displaystyle \left\{\begin{array}{l}

\\

\vare\,\varphi - (\imp\cdot\sigbf)\,\chi \ug m\,\varphi\\

\\

\vare\,\chi - (\imp\cdot\sigbf)\,\varphi \ug - m\,\chi\,,\\

\end{array}\right.}
\ee
and from the latter one:
\bb
{\dis\left\{\begin{array}{l}

\\

-\vare\,\varphi + (\imp\cdot\sigbf)\,\chi \ug m\,\varphi\\

\\

-\vare\,\chi + (\imp\cdot\sigbf)\,\varphi \ug - m\,\chi\,.\\

\end{array}\right.}
\ee
From system (27) we deduce explicitly the positive-energy amplitudes and from
(28) the negative-energy ones:
\bb
u_p \ug
\left(\begin{array}{cc}
\varphi\\
\ \\
{{\imp\cdot\sigbf}\over{\vare \, + \, m}}\,\varphi
\end{array}\right)\,,
\qquad\qquad
v_p \ug
\left(\begin{array}{cc}
{{\imp\cdot\sigbf}\over{\vare \, + \, m}}\,\chi\\
\ \\
\chi
\end{array}\right)\,.
\ee
For the helicity eigenfunctions, as before, we require the normalization
\ $\psid\psi=1/V$, which now implies the usual constraint for the amplitudes
\ $\upl^\dg\upl \ug \vpl^\dg\vpl \ug 2\vare$.
\ Consequently, we have [now $k=\sqrt{(\vare +m)(\vare -m)}\,$]:
\bb
\upl \ug \sqrt{\vare+m}
\left(\begin{array}{cc}
\teta\\
\ \\
{{\la k}\over{\vare + m}}\,\teta
\end{array}\right)\,,
\qquad \qquad
\vpl \ug \sqrt{\vare+m}
\left(\begin{array}{cc}
{{-\la k}\over{\vare + m}}\,\tetam\\
\ \\
\tetam
\end{array}\right)\,.
\ee

\

\noindent Let us come back to the PT wave-mechanics and to the solutions of PT Dirac
equation.

The limiting-case of massless ($p^2=0$) particles ---often called {\em
luxons}$^{[11]}$--- is immediately obtained,
from solutions (22),
by assuming $m=0; \ k=\vare$. \ In this way we obtain the two
``chiral'' amplitudes ($\gc$-eigenstates) $w_R$, $w_L$:
\bb
u_{p,+1} \ug v_{p,-1} \equiv w_R \ug \sqrt{k}\,
\left(\begin{array}{cc}
\theta_1\\
\ \\
\theta_1
\end{array}\right)\,,
\ee
\bb
u_{p,-1} \ug -v_{p,+1} \equiv w_L \ug \sqrt{k}\,
\left(\begin{array}{cc}
\theta_{-1}\\
\ \\
-\theta_{-1}
\end{array}\right)\,.
\ee
As they must, such solutions are identical to the ones foreseen by the standard
Dirac equation (and describing spinning massless particles, endowed with
four distinct states if they are charged ``Dirac particles'', and with two
distinct states if they are uncharged ``Majorana particles'').

\

\noindent It is interesting, at this point, to consider the so-called {\em trascendent
tachyons}$^{[12]}$, which, by definition, are particles endowed with
nonzero mass, but zero total energy:
\bb
\vare \ug 0\,, \qquad k \ug m\,.
\ee
The trascendent {\em classical} tachyons travel at infinite speed
$V\equiv k/\vare^{\#1}$, representing therefore the ``dual'' case (for the definition
of ``duality'' see
the fourth section) of the non-relativistic case,
\ $v=c/V\longr0,\: \imp=0,\:
\vare=m$.
\footnotetext{$^{\#1}$We shall show in the fourth section that the trascendent {\em
quantum} tachyons, i.e. the trascendent PT's,
behave quite differently.}
For this reason it is not surprising that, in the same way as it happens for
bradyons in the center-of-mass frame, {\em for trascendent PT's the equations for
spinors $\varphi$ and $\chi$ decouple} as well.$^{\#2}$
\footnotetext{$^{\#2}$Of course, these two cases do differ in the phase wave:
\ $\erm^{-imt}$
\ for the non-relativistic case, and
\ $\erm^{i\imp\cdot\xbf}$
\ for trascendent PT's.}
As a matter of fact, making $\vare=0$ in system (27) we have for the bispinorial
components of $u_p$:
\bb
{\dis\left\{\begin{array}{l}

\\

(\imp\cdot\sigbf \, + \, m)\,\chi \ug 0\\

\\

(\imp\cdot\sigbf \, - \, m)\,\varphi \ug 0\,.\\

\end{array}\right.}
\ee
(here we do not consider $v_p$, because the distinction between $u_p$ and $v_p$
is merely formal and concerns exclusively the sign of $\imp$ and not the one of
$\vare$, which is zero).
\ For the trascendent polarized states ($\lao$-eigenstates), since $k=m$ and then
$(\imp\cdot\sigbf)\,u_\la =m\la\,u_\la$, we shall have
\bb
{\dis\left\{\begin{array}{l}

\\

(\la\, + \,1)\,\chi \ug 0\\

\\

(\la\, - \,1)\,\varphi \ug 0\,;\\

\end{array}\right.}
\ee
and then:
\bb
u_{\la=+1} \ug
\left(\begin{array}{cc}\varphi\\ 0\end{array}\right)
\ug \sqrt{2k}\,
\left(\begin{array}{cc}\theta_{+1}\\ 0\end{array}\right)\,,
\ee
\bb
u_{\la=-1} \ug
\left(\begin{array}{cc}0\\ \chi\end{array}\right)
\ug \sqrt{2k}\,
\left(\begin{array}{cc}0\\ \theta_{-1}\end{array}\right)\,.
\ee

\

\

\section{$C$, $P$ and $T$ transformations and relativistic covariance}

We are now going to study the behaviour of our theory under the main discrete
symmetry operations, space-inversion $P$,\ charge-conjugation $C$,\
time-inversion $T$,\ 4-inversion $I$,\ and under the Lorentz homogeneous
proper group $L$.
We shall see that {\em the PT theory is symmetric under all the above
transformations; the operators related to the transformations $T$,
$I$ and $L$ are the same for both bradyons and PT's, whilst the operators
related to $P$ and $C$ result multiplied by $\gc$ when passing from the bradyonic to the PT
sector}. The procedure we shall follow to determine the symmetry operators is
the one usually employed in the standard Dirac theoretical framework.

\

\hskip 1cm PARITY

\noindent Starting from the PT Dirac equation (3)
\bb
(i\go\pa_t - i\vecna\cdot\gabf\,-\,m\gc)\,\psi(t; \xbf) \ug 0\,,
\ee
we write the equation for the space-inverted wave-function:
\bb
(i\go\pa_t + i\vecna\cdot\gabf\,-\,m\gc)\,U_P\,\psi(t; -\xbf) \ug 0\,,
\ee
quantity $U_P$ being a suitable {\em unitary} matrix operator.
If we want P-invariance we must demand that, after having applied a unitary
transformation to eq.~$\!$(39), $\psi(t; -\xbf)$ obey, as $\psi(t; \xbf)$, the original
eq.~$\!$(38). In order that this requirement be satisfied a matrix $U_P$ must exist
such that:
\bb
(i\go\pa_t + i\vecna\cdot\gabf\,-\,m\gc)\,U_P \ug \pm
U_P\,(i\go\pa_t - i\vecna\cdot\gabf\,-\,m\gc)\,.
\ee
In fact, this is equivalent to impose that the l.h.s. of the space-inverted
equation (39), after the unitary transformation $U_P^{-1}$, become equal, apart
from an unessential
sign, to the l.h.s. of equation (38):
\bb
U_P^{-1}\,(i\go\pa_t + i\vecna\cdot\gabf\,-\,m\gc)\,U_P \ug \pm
\,(i\go\pa_t - i\vecna\cdot\gabf\,-\,m\gc)\,.
\ee
If we take in eq.~$\!$(40) the plus sign before $U_P$, an operator satisfying the
above constraint does not exist; on the contrary, following the other choice
the searched matrix results to be $\erm^{i\al}\go\gc$. Quantity $\erm^{i\al}$ is
an unessential
phase factor because its choice does not affect the unitary
character of the operator.$^{\#3}$
 \footnotetext{$^{\#3}$As a matter of fact, if $O\longr\erm^{i\al}O$ from
 $OO^{-1}=\1$ it follows that $O^{-1}\longr\erm^{-i\al}O^{-1}$; at the same time
 $O^\dg\longr\erm^{-i\al}O^{\dg}$.
 Therefore, since both $O^{-1}$ and $O^\dg$ suffer the same transformation,
 their equivalence, $O^{-1}=O^\dg$, continues to hold.}
We may choose $\al=0$, so that:
\bb
U_P \ug \go\gc\,.
\ee
As is well-known$^{[13]}$, for bradyons analogous reasonings lead to the Dirac
parity operator that is formally different from (42), namely:
\ $U_P=\go$.
As for the bradyonic sector, also for the PT sector quantity
$\,m\,U_P$ coincides
with the ``mass term'' in the hamiltonian, \ $\Ho = -i\albf\cdot\vecna + m\go\gc$. \
As well as in the bradyonic theory, for the PT hamiltonian we have the parity
invariance, so that the PT energy is a space-inversion scalar:
\bb
U_P\,\Ho(-\xbf)\,U_P^{-1} \ug (\go\gc)\,[-i\albf\cdot(-\vecna) + m\go\gc]\,
(-\go\gc) \ug \Ho(\xbf)\,.
\ee
Furthermore, we may easily see that in the PT lagrangian (1) both the terms are
pseudoscalar quantities, even the mass term $m\psib\psi$, which instead in
the Dirac theory actually constitutes a scalar term. Thus, having a definite parity,
our lagrangian is parity-conserving and the consequent theory is invariant under
space-inversion.

\

\hskip 1cm CHARGE-CONJUGATION

\noindent The PT Dirac equation in an external field $A^\mu$ writes:
\bb
[\gm(i\pam - qA_\mu)\,-\,m\gc]\,\psi \ug 0\,.
\ee
Performing the complex conjugation and reversing all the signs, we obtain
\bb
[{\gm}^{\st}(i\pam + qA_\mu)\,+\,m{\gc}^{\st}]\,\psis \ug 0\,.
\ee
The $C$-transformed wave-function is, by definition, obtainable through application
of a suitable unitary operator $U_C$ on $\psis$: \ $\psi_C\equiv U_C\psis$.
For the $C$-invariance of the theory to hold we now have to demand that $U_C$
be such that $U_C\psis$ obey the original eq.~$\!$(44), but with opposite charge +$q$:
\bb
[{\gm}^{\st}(i\pam + qA_\mu)\,+\,m{\gc}^{\st}]\,U_C \ug
\pm U_C\,[\gm(i\pam + qA_\mu)\,-\,m\gc]\,.
\ee
Since we have (at least in the usual representations)
\bb
({\ga^{0,1,3,5}})^{\st}\ug\ga^{0,1,3,5}\,, \qquad
{\ga^2}^{\st}\ug-\ga^2\,,
\ee
the only possible choice (the minus sign before $U_C$) is the following, upto
phase factors:
\bb
U_C \ug i\,\ga^2\gc\,.
\ee
As for the parity operator, we see that the PT charge-conjugation operator is
obtained from the Dirac homologous operator times $\gc$.

\

\hskip 1cm TIME-INVERSION

\noindent Always in analogy with the standard theory, we write down the
``time-inverted'' PT Dirac equation:
\bb
[-i\go\pa_t - i\gabf\cdot\vecna\,-\,m\gc]^{\st}\,U_T\psis(-t; \xbf) \ug 0\,,
\ee
or, performing the complex conjugation,
\bb
[i{\go}^{\st}\pa_t + i{\gabf}^{\st}\cdot\vecna\,-\,m{\gc}^{\st}]\,U_T\psis(-t;
\xbf) \ug 0\,.
\ee
In order that the time-inversion invariance holds we must require:
\bb
[i{\go}^{\st}\pa_t + i{\gabf}^{\st}\cdot\vecna\,-\,m{\gc}^{\st}]\,U_T \ug
\pm U_T\,[i\go\pa_t - i\gabf\cdot\vecna\,-\,m\gc]\,.
\ee
Taking the plus sign we shall have for the time-inversion matrix:
\bb
U_T \ug i\,\ga^1\ga^3\,,
\ee
which is identical to the homologous Dirac operator.

\newpage

\hskip 1cm  4-INVERSION  AND $PCT$ TRANSFORMATIONS

\noindent The request for symmetry under 4-inversion,
i.e.
after $(t, \xbf)\longr(-t;
-\xbf)$, \ is a consequence of the relativistic invariance because
4-inversion is equivalent to a suitable proper Lorentz transformation.
The PT equation for the spacetime-inverted wave-function writes:
\bb
(-i\gm\pam\,-\,m\gc)\,U_{I}\psi(-t; -\xbf) \ug 0\,.
\ee
In order to have symmetry under 4-inversion we must find an unitary $U_I$ such that:
\bb
(-i\gm\pam\,-\,m\gc)\,U_{I} \ug \pm U_{I}\,(i\gm\pam\,-\,m\gc)\,.
\ee
The plus sign yields a PT 4-inversion operator identical to the bradyonic one:
\bb
U_I \ug i\gc\,.
\ee
It is easy to see, by a direct evaluation, that we have:
\bb
U_PU_CU_T \ug U_I\,, \qquad PCT\psi(t; \xbf) \ug
I\psi(t; \xbf) \ug i\gc\psi(-t; -\xbf)\,.
\ee
We have therefore found that the present theory is symmetric under spacetime
inversion and that the $PCT$ and 4-inversion transformations are equivalent
symmetry operations.
In such a way the invariance under $PCT$ descends from the relativistic
invariance itself.

\

\hskip 1cm RELATIVISTIC COVARIANCE

\noindent With regard to the relativistic covariance of the Dirac equation, we may
read in ref.~$\!$[14]: {\em ``...this form invariance of the Dirac equation
expresses the Lorentz invariance of the underlying energy-impulse
connection $p_\mu p^\mu=m^2$...''}. In the same way, the Lorentz ($L$)
covariance
of the PT Dirac equation (3) ---which we are going to prove--- expresses the
relativistic invariance of the PT kinematical constraint $p^2=-m^2$.

The proper homogeneous $L$
group generates the following transformations on spacetime coordinates:
\bb
(x^\nu)^{'} \ug {a^{\nu}}_\mu x^\mu\,,
\ee
that, in a compact form, we may write as $x^{'}=ax$. \
The transformation tensor $a$ reads:
\bb
{a^{\nu}}_\mu \ug {g^{\nu}}_\mu + \De{\om^{\nu}}_\mu\,,
\ee
where the $\De{\om^{\nu}}_\mu$ are the six independent generators of the
infinitesimal $L$-transformations.
In the standard theory$^{[14]}$ the $L$-transformed Dirac equation writes
\bb
[i\gm\pam\,-\,m]\,S(a)\psi^{'}(x^{'}) \ug 0\,,
\ee
provided that the $L$-transformed
wave-function is obtained out from the initial
one through the application of the linear (in general not unitary) matrix
$S^{-1}(a)$
\bb
\psi(x) \ug S(a)\,\psi^{'}(x^{'})\,, \qquad\qquad
\psi^{'}(x^{'}) \ug S^{-1}(a)\,\psi(x)\,,
\ee
with $S(a^{-1})=S^{-1}(a)$. \
As a function of $x^{'}$ only, eq.~$\!$(59) becomes:
\bb
[i\gm {a^{\nu}}_\mu\panb\,-\,m]\,S(a)\psi^{'}(x^{'}) \ug 0\,,
\ee
If we want the $L$-covariance, we must impose that this transformed equation
be put in the canonic form of the Dirac equation \ $(i\gm\pamb - m)\psi(x^{'})=0$. \
Then, as before, we have to require:
\bb
\gm {a^{\nu}}_\mu\,S(a) \ug S(a)\,\ga^\nu\,.
\ee
This equation is satisfied by the following linear operator, where
$\sig_{\mu\nu}\equiv {i\over 2}\,(\ga_\mu\ga_\nu - \ga_\nu\ga_\mu)$ denotes the
antisymmetric ``spin tensor'',
\bb
S \ug \1 -\frac{i}{4}\,\sig_{\mu\nu}\De\om^{\mu\nu}\,.
\ee
Repeating now these considerations for the PT sector, we have for the PT
$L$-transformed equation:
\bb
[i\gm {a^{\nu}}_\mu\panb\,-\,m\gc]\,\Sb (a)\psi^{'}(x^{'}) \ug 0\,.
\ee
Demanding the relativistic covariance now leads to the two contemporary
conditions for the $\Sb$ matrix:
\bb
\gm {a^{\nu}}_\mu\,\Sb (a) \ug \Sb (a)\,\ga^\nu\,, \qquad \quad
\gc\,\Sb (a)  \ug  \Sb (a)\,\gc\,.
\ee
For the commutation of the spin tensor with $\gc$, we are induced to conclude
that {\em such an operator coincides with the standard one given by {\rm (63)},
$\,S=\Sb\,$, and that lagrangian {\rm (1)}, which originates our theory, is
relativistically invariant}.
Thus, also in the present PT theory, the matrices $\gm$ transform as components of a 4-vector, $\gc$ transforms
as a pseudoscalar quantity, the bilinear $\psib\gm\psi$ transforms as a (true)
4-vector, the bilinear $\psib\gc\gm\psi$ transforms as a (pseudo)vector, and so on.

\

\

\section{Duality between momentum and velocity}

Starting from the PT hamiltonian (8) we can write the Heisenberg equation for
the velocity operator:
\bb
\vbo \equiv \xbdo \ug i\,[\Ho,\ \xbf]\,;
\ee
the solution of this operatorial equation is identical to Dirac's:
\bb
\vbo \ug \albf\,.
\ee
As implied in the denomination itself, {\em classical} tachyons,$^{[12]}$
perform superluminal motions, so that they behave as ``true'' tachyons. By contrast,
we are going to show that {\em quantum spinning PT's perform subluminal mean
motions}, just as bradyons do.
For this reason we have proposed to call them {\em Pseudotachyons}.
Let us now calculate explicitly
\ $\vbm\equiv\frac{\int\psid\albf\,\psi\,\dV}{\int\psid\psi\,\dV}$, \
and, in so doing, let us
continue to confront the present PT theory with the bradyonic one.
Remind that, like it occurs in the Dirac theory, {\em also PT hamiltonian}
(8) {\em do not commute with the velocity operator $\albf$, so that $\vbf$,
differently from $\imp$, is not determined and conserved}.
For simplicity we consider plane waves,
normalized, as anticipated in the second section, ``to one particle in the the considered
volume $V$'':
\bb
\int\psid\psi\,\dV = 1\,.
\ee
The impulse-helicity eigenstates for ordinary Dirac particles write
\bb
\psi^{(+)}_{p\la} \equiv \nbr\,\upl\erm^{-ipx}\,, \qquad \qquad
\psi^{(-)}_{p\la} \equiv \nbr\,\vpl\erm^{ipx}\,,
\ee
where the amplitudes $\upl, \vpl$ are given by eqs.~$\!$(30).
For the average velocity, through eqs.~$\!$(30),
we obtain the well-known momentum-velocity relation of classical
mechanics ($\vc$ denoting the classical expression of the velocity)
\bb
\vbm
\ug \int\psid_{p\la}\albf\,\psi_{p\la}\,dV
\equiv V\,\psid_{p\la}\albf\,\psi_{p\la} \ug
\frac{\imp}{\vare} \ug \vc\,,
\ee
for both the positive-energy and the negative-energy states.

Analogous considerations and evaluations can be made for the PT sector.
We have
\bb
\psi^{(+)}_{p\la} \equiv \npt\,\upl\erm^{-ipx}\,, \qquad \qquad
\psi^{(-)}_{p\la} \equiv \npt\,\vpl\erm^{ipx}\,,
\ee
with the amplitudes $\upl, \vpl$ given by eqs.~$\!$(22). For the average velocity,
through eqs.~$\!$(22), we now find that {\em the average velocity is
equal to the so-called ``dual velocity''$^{[12]}$ of the classical velocity}
$\vc^{\#4}$:
 \footnotetext{$^{\#4}$Let us recall that, in {\em Extended Relativity},$^{[12]}$
 a theory {\em not} invoked in the present work, a Lorentz transformation
 associated to a boost $V>1$ is obtained by the composition of a Lorentz
 transformation with boost $v={1\over V}$ dual of $V$, and of a trascendent
 Lorentz transformation, associated to an infinite-velocity ``boost''. In
 other words, in Extended Relativity, the relative velocity of two bodies
 endowed with reciprocally dual velocities is infinite, and {\em viceversa}.}
\bb
\vbm
\ug \int\psid_{p\la}\albf\,\psi_{p\la}\,dV
\equiv V\,\psid_{p\la}\albf\,\psi_{p\la} \ug
\frac{\vare\,\imp}{k^2}\,, \qquad |\vbm| \ug {1\over {|\vc|}}\,.
\ee
for both the positive-energy and the negative-energy states.
Since for PT's it is $|\vc|>1$, we find that {\em for PT's the average speed
is smaller than the light speed}.

\

\

\noindent Starting from the above solutions of the standard Dirac and PT Dirac equations, we
now work out the costraints linking 4-impulse with mass, with mean 4-velocity $\vmm$
and with mean spin 4-vector (or polarization 4-vector), $\smm$, for
bradyons and PT's, respectively.

Let us introduce the 4-vector $\pmut^\mu$, which may be
defined as {\em dual} of the 4-vector $p^\mu \equiv (\vare; \,\imp)$:
\bb
\pmut^\mu = \left(k; \ {{\vare\imp}\over {k}}\right)\,,
\ee
with $p\pmut = 0$.
Quantity $p^\mu$ is a timelike 4-vector for bradyons and a spacelike 4-vector
for PT's, respectively.
By contrast, {\em quantity $\pmut^\mu$ is a spacelike {\rm 4}-vector for bradyons and
a timelike {\rm 4}-vector for PT's}: in fact $\pmut^2=k^2-\vare^2=-p^2$. \
As a consequence, $k$ is the time component of a timelike 4-vector [eq.~$\!$(73)] for
PT's, while $\vare$ is the time component of a timelike 4-vector
($p^\mu$) for bradyons. As is well-known, the product $\vare V$ constitutes
a Lorentz-scalar quantity for ordinary particles. Analogously, for the just said
considerations, {\em the product $kV$ actually constitutes a Lorentz-invariant
quantity for PT's}.

Because \ $\vbm = \int\psid\albf\psi\,dV\equiv\int\psib\gabf\psi\,dV$ \ and
\ $\sbm = \int\psid\albf\gc\psi\,dV\equiv\int\psib\gabf\gc\psi\,dV$, \
we may define, as usual,
the following 4-vectorial quantities {\em referring to
bradyons}:
\bb
\vmm \equiv \frac{\vare }{m}\,\int\psib\gm\psi\,dV
\ug \frac{\vare V}{m}\,\psib\gm\psi
\ug {1\over m}\,(\vare; \ \imp)
\ug {{p^\mu}\over m}\,,
\ee
and
\bb
\smm \equiv \frac{\vare }{m}\,\int\psib\gm\gc\psi\,dV
\ug \frac{\vare V}{m}\,\psib\gm\gc\psi
\ug {\la\over m}\,\left(k; \ \frac{\vare\imp}{k}\right)
\ug \la\,{{\pmut^\mu}\over m}
\ee
[as it is found employing eqs.~$\!$(30,~$\!$69)]. We have $\vm^2=1$ and $\sm^2=-1$,
and in the ``quiet frame'', \ $\vbm=0,\ k=0,\ \som=0,\ |\sbm|=1$,\ as expected.

{\em For PT's}, analogously:
\bb
\vmm \equiv \frac{k}{m}\,\int\psib\gm\psi\,dV
\ug \frac{kV}{m}\,\psib\gm\psi
\ug {1\over m}\,\left(k; \ \frac{\vare\imp}{k}\right)
\ug {{\pmut^\mu}\over m}\,,
\ee
and
\bb
\smm \equiv \frac{k}{m}\,\int\psib\gm\gc\psi\,dV
\ug \frac{kV}{m}\,\psib\gm\gc\psi
\ug {\la\over m}\,(\vare; \ \imp)
\ug \la\,{{p^\mu}\over m}
\ee
[as it is found employing eqs.~$\!$(22,~$\!$71)].
As before we have a timelike 4-velocity ($\vm^2=1$) and a spacelike
4-polarization ($\sm^2=-1$), and in the ``quiet frame'',\
$\vbm=0,\ \vare=0,\ \som=0,\ |\sbm|=1$.

From the above equations we obtain the following constraints,
which are constitutive of the bradyonic kinematics:
\bb
{\dis\left\{\begin{array}{l}

\\

p^2 \ug m^2\\

\\

p_\mu \vmm \ug m\\

\\

p_\mu \smm \ug 0\,.\\

\end{array}\right.}
\ee
The PT kinematics is instead governed by the following rules:
\bb
{\dis\left\{\begin{array}{l}

\\

p^2 \ug -\,m^2\\

\\

p_\mu \vmm \ug 0\\

\\

p_\mu \smm \ug -m\la\,.\\

\end{array}\right.}
\ee
\noindent Notice the orthogonality between 4-impulse and 4-velocity which is a typical
property also of the kinematics of massless spinning particles.

\

\

\noindent Momentum and velocity are independent variables,
as expected for spinning particles which, having an intrinsic angular momentum,
are endowed with ``internal'' degrees of freedom.
In quantum theories for spin-$\um$ particles, in fact, the momentum operator,
$-i\vecna\;$, which conserves, and the velocity operator, $\albf\;$, which
does not (and whose components $\al_i$ do not commute), are independent
operators.
\ For Dirac particles the independence between the momentum and velocity
operators is at the origin ---for packets composed by opposite energies, generally
describing spatially-localized systems--- of the so-called {\em
zitterbewegung.}$^{\#5}$
  \footnotetext{$^{\#5}$The average velocity in the Dirac theory,\
  $\int\psid\albf\,\psi\,\dV$,\ for the most general
  wave-packet shows,$^{[13,15]}$ beyond the classical-like translational term
  $\imp\,/\vare$, the so-called zitterbewegung term involving a rapidly variating motion,
  with frequency $\omega=2\vare\geq 2m$. The zitterbewegung term has several observable consequences,
  of which the most famous is perhaps the {\em Darwin term} in the non-relativistic
  approximation of the Dirac hamiltonian.
  In a recent paper of the author$^{[16]}$ it is suggested that even the so-called
  ``quantum potential'', appearing in the hydrodynamical formulations of the
  Schr\"odinger and Pauli equations for electron, may be a very zitterbewegung
  effect.}
\ For PT's, as already seen, we have the fundamental consequence of the momentum-velocity
duality.

\

\

The duality between the dispersion velocity $\pa\vare/\pa\imp$ and the
travelling velocity of momentum and energy was found, even if under very
particular conditions, by Maccarrone and Recami $^{[17]}$ for {\em classical},
non-quantum tachyons.
Naranan$^{[18]}$ also obtained that for tachyons the observed velocity is dual
of the classical one, always inside a non-quantum context, by recourse to a simple generalization of
the ordinary Lorentz transformations by using complex variables.
In ref.~$\!$[19] Gott III showed that the motion of a Schwarzschild black hole,
associated with a spacelike world-line, does not involve transport of energy or
signals at speeds larger than the light speed.
Bathia and Pande$^{[20]}$ proved that for a degenerate gas of classical tachyons
the sound velocity is always subluminal.
In ref.~$\!$[21] Robinett found that the group speed of a spinless quantum tachyon
does not exceed the light speed, so that the Klein--Gorgon field
propagates at subluminal velocities; analogous results had been precedently
found by Fox {\em et al.}$^{[22]}$, Ecker$^{[9]}$ and Bers {\em et al.}$^{[23]}$,
and have been subsequently re-obtained by Ziolkowski$^{[24]}$.
[$\,$Let us notice, incidentally, that also for the electromagnetic radiation we may find
a kind of duality between superluminal group velocity and subluminal
transmission velocity of signals or energy-impulse. See in particular Strnad and
Kodre$^{[25]}$ who studied the transmission of electromagnetic waves across an
optically less-dense layer, that is, the propagation in a medium with anomalous
refraction index$\,$].
Let us recall that the spontaneous breaking of the gauge symmetries endowes
Higgs bosons (which are spinless particles with $m^2<0$) with a positive
mass squared. In such a way they may be eventually observed only like ordinary bradyonic
scalars.

From all these considerations we see that the momentum-velocity duality
has already been met in literature for scalars particles. By contrast, as far
as we know, {\em this property is quite a novelty for spin-$\um$ particles}.
Furthermore, for quantum scalar tachyons we have to consider the signals
transmission speed for very particular wave-packets, or in the presence
of special boundary conditions.
For PT's, instead, the momentum-velocity duality is intrinsic in the
theory, in that it is due to the presence of spin, and to the consequent
independence reciprocal between the momentum and velocity operators.
Anyway, either for spinless or spinning particles, we might assert that
Nature is always able to associate subluminal behaviours to negative masses
squared, in such a way hindering the observation of
charges and fields travelling at speeds greater than $c$.

\

\

\noindent The negative mass squared suggests that {\em the present theory might actually
describe (massive) neutrinos}.
From many recent experiments neutrinos seem to be endowed with a small non-zero
mass.$^{\#6}$
Furthermore, {\em in many laboratory experiments measuring direct kinematical limits,
the measured {\rm 4}-impulse
squared of neutrinos appears to be negative}. The
experimental data involving $p^2<0$ for electron [c)] and muon [a) and b)]
neutrinos has been deduced out from: a) observation of pion decay,$^{[28]}$ b)
precision measurements of $\pi^-$ mass,$^{[29]}$ c) kinematical analysis of
tritium beta-decay.$^{[30-35]}$
Let us mention the experimental results for $m^2$ (eV)$^2$ found in c):
\ $(-65\pm 85\pm 65)^{[30]}$; \ $(-24\pm 48\pm 61)^{[31]}$;
\ $(-147\pm 68\pm 41)^{[32]}$; $(-39\pm 34\pm 15)^{[33]}$;
\ $(-72\pm 41\pm 30)^{[34]}$. \
 \footnotetext{$^{\#6}$ As pointed out in ref.~$\!$[26], many experimental
 indications of a non-zero neutrino mass may be found mostly in atmospheric physics
 and astrophysics. Evidences come in particular from the solar neutrino deficit,
 (cf., e.g., the GALLEX experiment)
 and from the observation, by means of the Cosmic Background Explorer (COBE), of
 density fluctuations in the cosmic microwave background accounting for the
 existence of hot components of the ``mixed'' dark matter. Other evidences come
 from the detection of antineutrinos cooling the neutron star associated with the
 supernova SN 1987A$^{[27]}$, and from the recent Liquid Scintillator Neutrino
 Detector experiment at the Los Alamos National Laboratories. }
The idea that neutrinos might be tachyons (in the ``classical'' sense of the
word) dates back the end of sixties, and has been sometimes re-proposed.$^{[2,36]}$
Therefore, this hypothesis is not an absolute novelty.
Nevertheless, whenever the mentioned authors spoke of ``tachyonic neutrinos'',
they always asserted or supposed that the condition $p^2<0$ implied
faster-than-light motions. For such particles, as well as it occurs for classical tachyons,
it would result necessary the extension of the standard relativity to
superluminal Lorentz transformations (i.e. to boosts $>c$), with recourse
to the already mentioned {\em Extended Relativity}.
All that is not necessary for PT's, which, in our theory, are always subluminal
bodies. Thus, we may restrict ourselves to employ just the ordinary
transformations allowed in special relativity, only bearing in mind that the
classical dispersion law $\vbf=\imp/\vare$ does not hold anymore.
For the previous computations,
the speed which is foreseen for neutrinos will be given by:
\bb
v \ug \frac{\vare}{k} \ug \frac{\vare}{\sqrt{\vare^2 + m^2}}\,.
\ee
This expression does differ from the one holding for bradyons
\bb
u \ug \frac{k}{\vare} \ug \frac{\sqrt{\vare^2 - m^2}}{\vare}\,,
\ee
which is smaller than $c$ too, and from the one holding for classical tachyons
\bb
w \ug \frac{k}{\vare} \ug \frac{\sqrt{\vare^2 + m^2}}{\vare}\,,
\ee
which is instead greater than the light speed. The curves corresponding
to these three dispersion laws are reported in figure for a direct comparison.
Anyway, for very light particles as neutrinos, since $m^2\ll\vare^2$,
eachone of the three above expressions entails speeds experimentally
undistinguishable from $c$. Notice that for $m\longr0$ the plots referring to
the bradyonic and PT speeds overlap, and that for $v \longr0$ the PT dispersion
law shows a newtonian-like trend: $v\sim \vare/m$.
{\em Suited, highly precise experiments, looking into the small-energy
region of the dispersion law, might clearly show if neutrinos are,
or not, PT's.}

\

\noindent In a work in preparation$^{[10]}$ we carefully study
the PT current for the general wave-packet and try to perform the second quantization of
the theory.
\ In so doing, we find that the PT field obeys the same fermionic anticommutation
rules of the Dirac field.

\begin{figure}
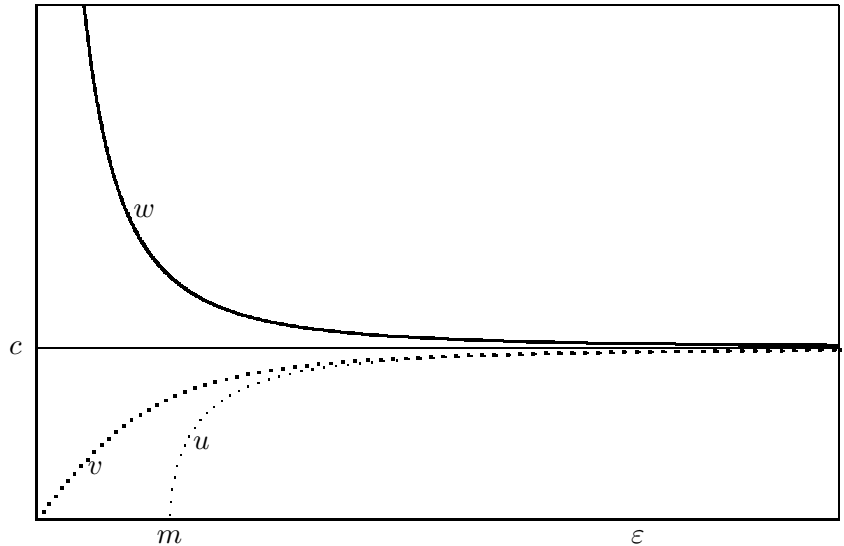

\begin{center}
% GNUPLOT: LaTeX picture
\setlength{\unitlength}{0.240900pt}
\ifx\plotpoint\undefined\newsavebox{\plotpoint}\fi
\sbox{\plotpoint}{\rule[-0.200pt]{0.400pt}{0.400pt}}%

\end{center}
\caption{Energy-speed dispersion plots for bradyons ($u$), \ PTs ($v$), \
and
classical tachyons ($w$)}
\end{figure}

\

\

{\bf Acknowledgements}

\noindent The author is very grateful to E. Recami for essential suggestions,
crucial discussions and for the constant cooperation. \
The kind collaboration of G. Andronico, G.G.N. Angilella, M. Carbonaro,
G. Di Lorenzo, S. Dimartino, P. Falsaperla, G. Fonte, G. Giuffrida, I. Inclimona,
S. Lo Nigro, G. Lucifora, D. Rapoport, M.T. Vasconselos, and in particular of
F. Raciti and L. Lo Monaco, is also acknowledged.

\

\newpage

\h {\large\bf Appendix}\\

\noindent {\large{\em Solutions of the PT Dirac equation in spinorial (Weyl's) representation}}\\

In the so-called {\em spinorial representation} (also said the {\em Weyl's
representation$\,$}) the matrices $\gm$, $\gc$ write:

\

$\hfill{\displaystyle
\go \equiv \left(
\begin{array}{cc}
\Orm   & \1\\
\1    &  \Orm\\
\end{array}
\right)\,,
\qquad
\gabf \equiv \left(
\begin{array}{cc}
\Orm     & - \sigbf\\
\sigbf   & \ \Orm\\
\end{array}
\right)\,,
\qquad
\gc \equiv \left(
\begin{array}{cc}
\1     &  \ \Orm\\
\Orm   &  - \1 \\
\end{array}
\right)\,.
}\hfill{\dis({\rm A.1})}$

\

\noindent In this representation the bispinorial amplitude in $\npt\,w\,\erm^{-ipx}$ may be
put in the following form:

\

$\hfill{\displaystyle
w \equiv
\left(\begin{array}{cc}\xi\\ \eta\end{array}\right)\,,
}\hfill{\dis({\rm A.2})}$

\

\noindent where $\xi, \eta$ are two-component {\em Weyl spinors} (``not pointed'' and ``pointed''
respectively).
The interrelation between the spinors in the present representation and the ones
in standard representation is:

\

$\hfill{\displaystyle
\varphi \equiv \frac{\xi\, + \,\eta}{\sqrt{2}}\,, \qquad \qquad
\chi \equiv \frac{\xi\, - \,\eta}{\sqrt{2}}\,.
}\hfill{\dis({\rm A.3})}$

\

\noindent The PT Dirac equation in spinorial representation writes:

\

$\hfill{\displaystyle
{\dis\left\{\begin{array}{l}

\\

(\vare + \imp\cdot\sigbf)\,\eta \ug m\,\xi\\

\\

(\vare - \imp\cdot\sigbf)\,\xi \ug -\,m\,\eta\\

\end{array}\right.}
}\hfill{\dis({\rm A.4})}$

\

\noindent for the positive-energy states, and

\

$\hfill{\displaystyle
{\dis\left\{\begin{array}{l}

\\

(-\vare - \imp\cdot\sigbf)\,\eta \ug m\,\xi\\

\\

(-\vare + \imp\cdot\sigbf)\,\xi \ug -\,m\,\eta\\

\end{array}\right.}
}\hfill{\dis({\rm A.5})}$

\

\noindent for the negative-energy ones. The relative solutions may be written
as follows:

\

$\hfill{\displaystyle
u_p \ug
\left(\begin{array}{cc}\xi\\
\ \\
{{\imp\cdot\sigbf\,-\,\vare}\over{m}}\,\xi
\end{array}\right)\,,
\qquad \qquad
v_p \ug
\left(\begin{array}{cc}-{{\imp\cdot\sigbf\, + \,\vare}\over{m}}\,\eta
\ \\
\eta
\end{array}\right)\,.
}\hfill{\dis({\rm A.6})}$

\

\noindent Bearing in mind the PT equality \ $m=\sqrt{(k+\vare\la)(k-\vare\la)}$, \ the
$\la$-eigenfunctions are:

\

$\hfill{\displaystyle
\upl \ug \sqrt{k + \vare\la}
\left(\begin{array}{cc}\teta\\
\ \\
{\la m\over{k + \vare\la}}\,\teta
\end{array}\right)\,,
\qquad\qquad
\vpl \ug \sqrt{k + \vare\la}
\left(\begin{array}{cc}
{m\over{k\la + \vare}}\,\tetam\\
\ \\
\la\tetam
\end{array}\right)\,.
}\hfill{\dis({\rm A.7})}$

\

\noindent These solutions are normalized, like in eqs.~$\!$(22), with $\psid\psi=1/V,
\ \upl^\dg\upl=\vpl^\dg\vpl=2k$.
As expected, also in spinorial representation a short calculation yields \
$\int\psid\albf\psi\,dV = \frac{\vare\imp}{k^2}$.

As previously made in standard representation, we write down, for comparison,
the solutions of the bradyonic Dirac equation in the Weyl representation.
We have for the bispinorial amplitudes:

\

$\hfill{\displaystyle
u_p \ug
\left(\begin{array}{cc}\xi\\
\ \\
{{\vare\,-\,\imp\cdot\sigbf}\over{m}}\,\xi
\end{array}\right)\,,
\qquad \qquad
v_p \ug
\left(\begin{array}{cc}-{{\vare\,+\,\imp\cdot\sigbf}\over{m}}\,\eta\\
\ \\
\eta
\end{array}\right)\,.
}\hfill{\dis({\rm A.8})}$

\

\noindent By using the bradyonic equality \ $m=\sqrt{(\vare+k\la)(\vare-k\la)}$, \ we have for
the helicity eigenfunctions (normalized, as in the standard representation, with
$\psid\psi=1/V, \ \upl^\dg\upl=\vpl^\dg\vpl=2\vare$):

\

$\hfill{\displaystyle
\upl \ug \sqrt{\vare + k\la}
\left(\begin{array}{cc}\teta\\
\ \\
{m\over{\vare + k\la}}\,\teta
\end{array}\right)\,,
\qquad  \qquad
\vpl \ug \sqrt{\vare + k\la}
\left(\begin{array}{cc}
{-m\over{\vare + k\la}}\,\tetam\\
\ \\
\tetam
\end{array}\right)\,.
}\hfill{\dis({\rm A.9})}$

\

\

\end{document}